\documentclass{iopart}                    % onecolumn
\usepackage{graphicx}
\begin{document}

\title{Bell's Experiment in Quantum Mechanics and Classical Physics}

\author{Tom Rother}

\address{German Aerospace Center, Remote Sensing Technology Inst., 
Kalkhorstweg 53, D-17235 Neustrelitz, GERMANY}
\ead{tom.rother@dlr.de}

\begin{abstract}
Both the quantum mechanical and classical Bell's experiment are within 
the focus of this paper. The fact that one measures different probabilities 
in both experiments is traced back to the superposition of two orthogonal 
but non-entangled substates in the quantum mechanical case. This superposition 
results in an interference term that can be splitted into two additional states 
representing a sink and a source of probabilities in the classical event space 
related to Bell's experiment. As a consequence, a statistical operator can 
be related to the quantum mechanical Bell's experiment that contains already 
negative quasi probabilities, as usually known from quantum optics in 
conjunction with the Glauber-Sudarshan equation. It is proven that the 
existence of such negative quasi probabilities are neither a sufficient nor a 
necessary condition for entanglement. The equivalence of using an interaction 
picture in a fixed basis or of employing a change of basis to describe Bell's 
experiment is demonstrated afterwards. The discussion at the end of this paper 
regarding the application of the complementarity principle to the quantum 
mechanical Bell's experiment is supported by very recent double slit experiments 
performed with polarization entangled photons.
\end{abstract}

\maketitle

\section{Introduction}
\label{sec:1}
Even if existing since nearly 100 years there are still controversial 
discussions regarding the epistemological consequences of quantum mechanics. 
We are obviously able to describe the behaviour of objects on the 
atomic and subatomic level in a quite formal mathematical way rather than 
to align it with our experience from classical physics. In the younger days of 
quantum mechanics those contradictions have been discussed on a pure philosophical 
level. But since the beginning of the 1980'th there exist several experimental 
results (and the number of corresponding experiments is growing continously even 
in our days) which seem to confirm the correctness of the strange behaviour of 
quantum objects. Two experiments are within the focus of these discussions. These 
are the famous double slit experiment, and Bell's experiment. The latter is usually 
considered to be an essential indication of the non-local character of quantum 
mechanics, and, strongly related to this, as an evidence of the existence of 
''entangled states''. It has its roots in the basic discussion regarding the 
completeness of quantum mechanics. This discussion was initiated by Einstein, 
Podolsky and Rosen (EPR) on the one side, and by Bohr on the other side in two 
famous papers published in 1935 (see \cite{journal1},\cite{journal2}). In the paper 
of EPR quantum mechanics was accused of being incomplete, and, therefore, that one 
has to look for hidden parameters to replace it by a complete theory. In his answer, 
Bohr defended his position of understanding quantum mechanics as a complete theory 
and his insistence on the principle of complementarity. But again, this discussion 
was a pure philosophical ones until the famous paper of Bell \cite{book1}. He 
derived therein an inequality (now called Bell's inequality) that allows for an 
experimental proof of the non-local character of quantum mechanics as well as the 
existence of entangled states. But it took again more than one decade until the 
first experiments with polarization entangeld photons provided us with an indication 
that Bell's inequality can indeed be violated in quantum mechanics. These 
experiments have been performed by A. Aspect and co-workers at the beginning of the 
1980'th \cite{journal3}. The existence of entangled states is the most essential 
difference between quantum mechanics and classical physics, according to Schr\"odinger. 
He wrote in \cite{journal4}: {\it When two systems, of which we know the states by 
their respective representatives, enter into temporary physical interaction due to 
known forces between them, and when after a time of mutual influence the system 
separate again, then they can no longer be described in the same way as before viz. 
by endowing each of them with a representative of its own. I would not call that one 
but rather that characteristic trait of Quantum mechanics, the one that enforces its 
entire departure from classical lines of thought. By the interaction the two 
representatives have become entangled}. 

Today, Schr\"odinger's position as well as the assumption that entanglement is 
responsible for the violation of Bell's inequality in quantum mechanics is well 
accepted among most of the physicists. However, it will be demonstrated in what 
follows that it is not primarily the entanglement but the superposition of 2 
non-entangled and orthogonal substates which do not belong to orthogonal subspaces 
that results in a violation of Bell's inequality. I.e., it is a similar reason we 
know already from the double slit experiment. Beside the well-known description of 
Bell's experiment in terms of a basis transformation it is shown moreover that 
these 2 substates are considered to be the result of 2 additional but local and 
stochastically independent interactions. This avoids the assumption of any 
''spooky action at a distance''. In this context it is quite interesting to see 
that we are able to relate a statistical operator already to both the quantum 
mechanical and classical Bell's experiment. The necessity of taking interference 
terms into account results in negative weights of the statistical operator related 
to the quantum mechanical Bell's experiment. Such negative weights are known so far 
only in quantum optics as negative ''quasi probabilities''. They are proven 
afterwards to represent neither a sufficient nor a necessary condition for 
entanglement. The statistical operator that belongs to a fixed parameter 
configuration of Bell's experiment can be simply extended to a ''Bell's ensemble'', 
i.e. to an incoherent mixture of different experimental configurations, as usually 
known from quantum mechanics. Some consequences especially with respect to the 
complementarity principle are finally discussed. 

\section{The quantum mechanical Bell's experiment}
\label{sec:2}
\subsection{The quantum mechanical Bell's box}
\label{subsec:1}
%%%%%%%%%%%%%%%%%%%%%%%%%%%%%%%%%%%%%%%%%%%%%%%%%%%%%%%%%%%%%%%%%%%%%
\begin{figure}
\centering
\includegraphics[height=6.cm]{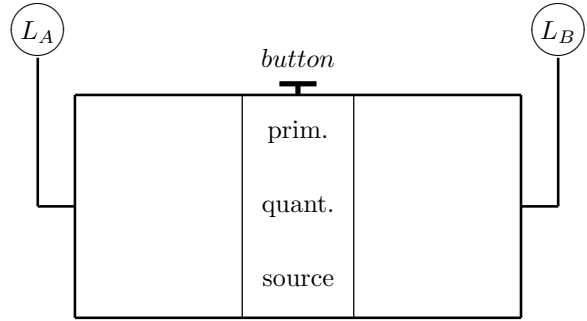}
\caption{Bell's box with only a primary quantum mechanical source}
\label{Abbildung_4_1}
\end{figure}
%%%%%%%%%%%%%%%%%%%%%%%%%%%%%%%%%%%%%%%%%%%%%%%%%%%%%%%%%%%%%%%%%%%%
To commemorate the well-known results let's start with a quite 
phenomenological description of the quantum mechanical Bell's 
experiment by introducing a ''quantum mechanical Bell's box'' (QBB). 
In its first level of configuration our QBB consists of a box with 3 
compartments (see Fig. \ref{Abbildung_4_1}). In the center compartment we 
have placed a source that emits 2 polarization entangled photons 
(horizontally (h)- and vertically (v)-polarized) into opposite directions 
once we push the button. But we don't know the state of polarization of the 
photon emitted in a certain direction. I.e., we don't know if we have the 
combination (h,v) or (v,h) with respect to the polarization of both photons 
in a single event. The first term in the 
brackets is related to the photon on the left hand side, and the 
second term is related to the photon on the right hand side. For the 
time being, the other 2 compartments on the left- and right hand side 
remain empty. 2 additional lamps $L_A$ and $L_B$ are mounted at the ends 
of the box. Each lamp is equipped with a detector that switches the lamp 
on if a h-polarized photon is detected. The lamp remains switched off 
otherwise. After performing a multitude of experiments with this first 
level setup of the QBB we realize that there are just two possible events 
(2 non-local measurement values). These are ''lamp $L_A$ on, lamp $L_B$ 
off: $(y_A,n_B)$'', and ''lamp $L_A$ off, lamp $L_B$ on: $(n_A,y_B)$''.
This defines our classical event space. Let's further assume that the 
primary stochastic source acts in such a way that the probability is $1/2$ 
for both measurement pairs. If we relate the eigenvector 
%%%%%%%%%%%%%%%%%%%%%%%%%%%%%%%%%%%%%%%%%%%%%%%%%%%%%%%%%%%%%%%%%%%%%
\begin{equation}
|\varphi_1> \, = \, (1,0)
\label{Def_1}
\end{equation}
%%%%%%%%%%%%%%%%%%%%%%%%%%%%%%%%%%%%%%%%%%%%%%%%%%%%%%%%%%%%%%%%%%%%%
to the measurement value ''lamp A/B on ($y_{A/B}$)'', and the 
eigenvector 
%%%%%%%%%%%%%%%%%%%%%%%%%%%%%%%%%%%%%%%%%%%%%%%%%%%%%%%%%%%%%%%%%%%%%
\begin{equation}
|\varphi_2> \, = \, (0,1)
\label{Def_2}
\end{equation}
%%%%%%%%%%%%%%%%%%%%%%%%%%%%%%%%%%%%%%%%%%%%%%%%%%%%%%%%%%%%%%%%%%%%%
to the measurement value ''lamp A/B off ($n_{A/B}$)'' then we are able 
to characterize this first level QBB by the probability state vector 
%%%%%%%%%%%%%%%%%%%%%%%%%%%%%%%%%%%%%%%%%%%%%%%%%%%%%%%%%%%%%%%%%%%%%
\begin{equation}
|\Phi_{QBB}^{(0)}> \, =  \, \frac{1}{\sqrt 2} \cdot \left( 
|\varphi_1,\varphi_2> \, - \; |\varphi_2,\varphi_1> \right)
\label{QBK_1}
\end{equation}
%%%%%%%%%%%%%%%%%%%%%%%%%%%%%%%%%%%%%%%%%%%%%%%%%%%%%%%%%%%%%%%%%%%%%
in the direct product space. Calculating the scalar product 
%%%%%%%%%%%%%%%%%%%%%%%%%%%%%%%%%%%%%%%%%%%%%%%%%%%%%%%%%%%%%%%%%%%%%
\begin{equation}
<\Phi_{QBB}^{(0)}|\Phi_{QBB}^{(0)}>
\label{QBK_1a}
\end{equation}
%%%%%%%%%%%%%%%%%%%%%%%%%%%%%%%%%%%%%%%%%%%%%%%%%%%%%%%%%%%%%%%%%%%%%
provides as with the probabilities $1/2$ for each pair $(y_A,n_B)$ and 
$(n_A,y_B)$ observed before in the experiments. 

%%%%%%%%%%%%%%%%%%%%%%%%%%%%%%%%%%%%%%%%%%%%%%%%%%%%%%%%%%%%%%%%%%%%%
\begin{figure}
\centering
\includegraphics[height=6.cm]{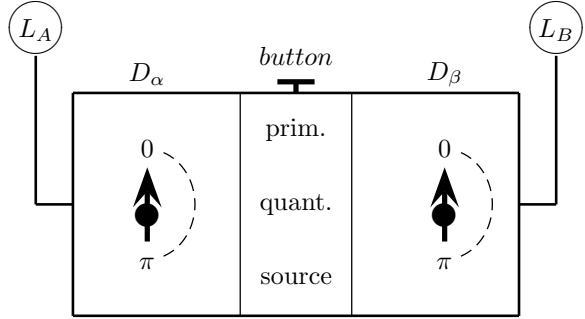}
\caption{Bell's box with the primary quantum mechanical source of 
Fig. \ref{Abbildung_4_1} but now with 2 additional, local stochastic 
interaction modules (polarization filters)}
\label{Abbildung_4_2}
\end{figure}
%%%%%%%%%%%%%%%%%%%%%%%%%%%%%%%%%%%%%%%%%%%%%%%%%%%%%%%%%%%%%%%%%%%%
In a second level of configuration we insert 2 additional stochastic 
interaction modules into the so far empty compartments of the QBB. 
These modules are nothing but 2 polarization filters the photon on each 
side is interacting with. The position of the polarization filters can 
be continuously varied between $[0,\pi]$ with corresponding rotating 
switches $D_\alpha$ and $D_\beta$. The local interactions become in 
this way functions of the local parameters $\alpha$ and $\beta$. Finally, 
the detectors of the lamps are substituted by new detectors which act in 
such a way that the lamps are switched on if a photon is detected, 
independent of its polarization. The lamps remain switched off otherwise. 
This second level configuration of our Bell's experiment is depicted in 
Fig. \ref{Abbildung_4_2}. It is now related to a 4-dim. event space 
defined by the 4 possible pairs $(y_A,y_B), (y_A,n_B), (n_A,y_B)$, and 
$(n_A,n_B)$. It is an essential advantage of Bell's experiment that it 
can be exclusively described in this 4-dim. space since 
there exists a condition to prove entanglement in a straightforward 
way. Performing a large number of experiments for different sets 
of the local parameters $\alpha$ and $\beta$ results in the following 
probability state in the 4-dim. direct product space that characterizes 
this second level QBB:
%%%%%%%%%%%%%%%%%%%%%%%%%%%%%%%%%%%%%%%%%%%%%%%%%%%%%%%%%%%%%%%%%%%%%
\begin{eqnarray}
|\Phi_{QBB}> \, =  \, c_{11} \cdot |\varphi_1,\varphi_1> + \, c_{12} 
\cdot |\varphi_1,\varphi_2> + ~~~~~~~~~~~~~~~~~~~
\nonumber\\
\, c_{21} \cdot |\varphi_2,\varphi_1> + \, c_{22} \cdot 
|\varphi_2,\varphi_2> \; ,
\label{ZZV_6}
\end{eqnarray}
%%%%%%%%%%%%%%%%%%%%%%%%%%%%%%%%%%%%%%%%%%%%%%%%%%%%%%%%%%%%%%%%%%%%%
with the $\alpha$ and $\beta$ dependent probability amplitudes
%%%%%%%%%%%%%%%%%%%%%%%%%%%%%%%%%%%%%%%%%%%%%%%%%%%%%%%%%%%%%%%%%%%%%
\begin{eqnarray}
c_{11} & = & c_{22} \, = \, \frac{1}{\sqrt 2} \cdot 
\sin (\alpha - \beta)
\label{ZZV_7}
\\
c_{21} & = & c_{12} \, = \, \frac{1}{\sqrt 2} \cdot 
\cos (\alpha - \beta) \; .
\label{ZZV_8}
\end{eqnarray}
%%%%%%%%%%%%%%%%%%%%%%%%%%%%%%%%%%%%%%%%%%%%%%%%%%%%%%%%%%%%%%%%%%%%%
The additional eigenvectors $|\varphi_1, \varphi_1>$ and 
$|\varphi_2, \varphi_2>$ are related to the 2 additional measurement 
pairs $(y_A,y_B)$ and $(n_A,n_B)$. If $\alpha=\beta$ is chosen, then 
we end up with (\ref{QBK_1}) as a special case. This probability 
state is normalized to unity, i.e.,
%%%%%%%%%%%%%%%%%%%%%%%%%%%%%%%%%%%%%%%%%%%%%%%%%%%%%%%%%%%%%%%%%%%%%
\begin{equation}
<\Phi_{QBB}|\Phi_{QBB}> \, =  \, 1 
\label{ZZV_9}
\end{equation}
%%%%%%%%%%%%%%%%%%%%%%%%%%%%%%%%%%%%%%%%%%%%%%%%%%%%%%%%%%%%%%%%%%%%%
holds. By the way, the same normalization holds obviously for the 
probability state vector (\ref{QBK_1}). It must again be emphasized 
that (\ref{ZZV_6}) with probability amplitudes (\ref{ZZV_7}) and 
(\ref{ZZV_8}) is considered to be the result of the measured 
probabilities
\begin{itemize}
\item probability $w(y,y)$/$w(n,n)$ that both lamps are switched 
on/switched off:
%%%%%%%%%%%%%%%%%%%%%%%%%%%%%%%%%%%%%%%%%%%%%%%%%%%%%%%%%%%%%%%%%%%%%
\begin{equation}
w(y,y) \, = \, w(n,n) \, =  \, \frac{1}{2} \cdot 
\sin^2(\alpha - \beta)
\label{ZZV_10}
\end{equation}
%%%%%%%%%%%%%%%%%%%%%%%%%%%%%%%%%%%%%%%%%%%%%%%%%%%%%%%%%%%%%%%%%%%%%
\item probability $w(y,n)$/$w(n,y)$ that just one lamp is switched 
on and the other lamp remains switched off:
%%%%%%%%%%%%%%%%%%%%%%%%%%%%%%%%%%%%%%%%%%%%%%%%%%%%%%%%%%%%%%%%%%%%%
\begin{equation}
w(n,y) \, = \, w(y,n) \, =  \, \frac{1}{2} \cdot 
\cos^2(\alpha - \beta)
\label{ZZV_11}
\end{equation}
%%%%%%%%%%%%%%%%%%%%%%%%%%%%%%%%%%%%%%%%%%%%%%%%%%%%%%%%%%%%%%%%%%%%%
\end{itemize}
These probabilities have been indeed observed in real experiments and 
can therefore accepted to be a fact. I would also like to point out that 
in the next subsection, when deriving the probability 
amplitudes (\ref{ZZV_7}) and (\ref{ZZV_8}) by use of a T-matrix, we 
will end up with relation $c_{21} = - c_{12}$ instead of relation 
$c_{21} = c_{12}$ in (\ref{ZZV_8}). But in this subsection we started from 
the phenomenological point of view that we have first measured the always 
positive probabilities. The related probability amplitudes are obtained 
afterwards by simply taking the square root of these probabilities. But 
this sign is of no importance when calculating the probabilities related 
to each single event from the scalar product of (\ref{ZZV_6}) with itself. 

Regarding Bell's experiment the most general probability state is 
obviously given by 
%%%%%%%%%%%%%%%%%%%%%%%%%%%%%%%%%%%%%%%%%%%%%%%%%%%%%%%%%%%%%%%%%%%%%
\begin{equation}
|\Phi> \, = \, \sum_{i,k=1}^2 \, c_{ik} \cdot |\varphi_i,\varphi_k> 
\label{Kap3_2_4}
\end{equation}
%%%%%%%%%%%%%%%%%%%%%%%%%%%%%%%%%%%%%%%%%%%%%%%%%%%%%%%%%%%%%%%%%%%%%
with
%%%%%%%%%%%%%%%%%%%%%%%%%%%%%%%%%%%%%%%%%%%%%%%%%%%%%%%%%%%%%%%%%%%%%
\begin{equation}
\sum_{i,k=1}^2 \, c^2_{ik} \, = \, 1 \; .
\label{Kap3_2_5}
\end{equation}
%%%%%%%%%%%%%%%%%%%%%%%%%%%%%%%%%%%%%%%%%%%%%%%%%%%%%%%%%%%%%%%%%%%%%
On the other hand, we have the 2 general substates
%%%%%%%%%%%%%%%%%%%%%%%%%%%%%%%%%%%%%%%%%%%%%%%%%%%%%%%%%%%%%%%%%%%%%
\begin{eqnarray}
|\Phi_l> & = & \sum_{i=1}^2 \, c_i \cdot 
|\varphi_i>
\label{Kap3_2_7a}
\\
|\Phi_r> & = & \sum_{i=1}^2 \, {\tilde c}_i \cdot |\varphi_i>
\label{Kap3_2_7b}
\end{eqnarray}
%%%%%%%%%%%%%%%%%%%%%%%%%%%%%%%%%%%%%%%%%%%%%%%%%%%%%%%%%%%%%%%%%%%%%
of the 2-dim. subspaces related to the events on the left- and right 
hand side of Bell's experiment. Their real valued probability 
amplitudes are also normalized to unity, 
%%%%%%%%%%%%%%%%%%%%%%%%%%%%%%%%%%%%%%%%%%%%%%%%%%%%%%%%%%%%%%%%%%%%%
\begin{eqnarray}
\sum_{i=1}^2 \, c^2_i  & = & 1
\label{Kap3_2_7c}
\\
\sum_{i=1}^2 \, {\tilde c}^2_i  & = & 1 \; .
\label{Kap3_2_7d}
\end{eqnarray}
%%%%%%%%%%%%%%%%%%%%%%%%%%%%%%%%%%%%%%%%%%%%%%%%%%%%%%%%%%%%%%%%%%%%%
The probability state (\ref{Kap3_2_4}) can be resolved into the 
direct product of the substates (\ref{Kap3_2_7a}) and (\ref{Kap3_2_7b}) 
if the real valued amplitude functions $c_{ik}$ are given by
%%%%%%%%%%%%%%%%%%%%%%%%%%%%%%%%%%%%%%%%%%%%%%%%%%%%%%%%%%%%%%%%%%%%%
\begin{equation}
c_{ik} \, = \, c_i \cdot {\tilde c}_k \; ,
\label{Kap3_2_9}
\end{equation}
%%%%%%%%%%%%%%%%%%%%%%%%%%%%%%%%%%%%%%%%%%%%%%%%%%%%%%%%%%%%%%%%%%%%%
and if condition
%%%%%%%%%%%%%%%%%%%%%%%%%%%%%%%%%%%%%%%%%%%%%%%%%%%%%%%%%%%%%%%%%%%%%
\begin{equation}
c_{11} \cdot c_{22} \, = \, c_{12} \cdot c_{21}
\label{Kap3_2_6}
\end{equation}
%%%%%%%%%%%%%%%%%%%%%%%%%%%%%%%%%%%%%%%%%%%%%%%%%%%%%%%%%%%%%%%%%%%%%
holds. The probability amplitudes of (\ref{Kap3_2_7a}) and 
(\ref{Kap3_2_7b}) can then be recalculated from the amplitudes of 
(\ref{Kap3_2_4}) according to 
%%%%%%%%%%%%%%%%%%%%%%%%%%%%%%%%%%%%%%%%%%%%%%%%%%%%%%%%%%%%%%%%%%%%%
\begin{eqnarray}
c_i^2 & = & \sum_{k=1}^2 \, c_{ik}^2 \quad ; \; i=1,2
\label{Kap3_2_8a}
\\
{\tilde c}_i^2 & = & \sum_{k=1}^2 \, c_{ki}^2  \quad ; \; i=1,2 \; .
\label{Kap3_2_8b}
\end{eqnarray}
%%%%%%%%%%%%%%%%%%%%%%%%%%%%%%%%%%%%%%%%%%%%%%%%%%%%%%%%%%%%%%%%%%%%%
However, if condition (\ref{Kap3_2_6}) is violated, we are unable to 
resolve (\ref{Kap3_2_4}) into the direct product of (\ref{Kap3_2_7a}) 
and (\ref{Kap3_2_7b}), and (\ref{Kap3_2_4}) is called ''entangled''.
Regarding probability state (\ref{ZZV_6}) with amplitudes 
(\ref{ZZV_7}) and (\ref{ZZV_8}) condition (\ref{Kap3_2_6}) holds 
only if
%%%%%%%%%%%%%%%%%%%%%%%%%%%%%%%%%%%%%%%%%%%%%%%%%%%%%%%%%%%%%%%%%%%%%
\begin{equation}
\sin^2(\alpha - \beta) \, =  \, \cos^2(\alpha - \beta) \; ,
\label{ZZV_14}
\end{equation}
%%%%%%%%%%%%%%%%%%%%%%%%%%%%%%%%%%%%%%%%%%%%%%%%%%%%%%%%%%%%%%%%%%%%%
i.e., if we have $\alpha-\beta=\pi/4$ for the corresponding local 
interaction parameters. Only then we are able to resolve this 
probability state into the direct product of the two 2-dim. subspaces. 
For all other parameter configurations $\alpha$ and $\beta$ this state 
becomes entangled. 

Once we know the probabilities of a certain experimental configuration 
(i.e., for a certain choice of the local parameters $\alpha$ and 
$\beta$) we are able to calculate the so-called ''correlation 
function'' $K(\alpha,\beta)$ according to
%%%%%%%%%%%%%%%%%%%%%%%%%%%%%%%%%%%%%%%%%%%%%%%%%%%%%%%%%%%%%%%%%%%%%
\begin{equation}
K(\alpha,\beta) \, =  \, w(y,y) \, + \, w(n,n) \, - \, w(y,n) 
\, - \, w(n,y) \, =  \, - \, \cos 2(\alpha - \beta) \; .
\label{ZZV_15}
\end{equation}
%%%%%%%%%%%%%%%%%%%%%%%%%%%%%%%%%%%%%%%%%%%%%%%%%%%%%%%%%%%%%%%%%%%%%
This function is the essential quantity in Bell's inequality. Looking 
into the relevant literature (see \cite{journal5}, for example) we 
find
%%%%%%%%%%%%%%%%%%%%%%%%%%%%%%%%%%%%%%%%%%%%%%%%%%%%%%%%%%%%%%%%%%%%%
\begin{equation}
| \, K(\alpha,\beta) \, - \, K(\alpha,\beta') \, | \, + \, 
| \, K(\alpha',\beta) \, + \, K(\alpha',\beta') \, | \, - \, 2 
\le \, 0 \; .
\label{ZZV_16}
\end{equation}
%%%%%%%%%%%%%%%%%%%%%%%%%%%%%%%%%%%%%%%%%%%%%%%%%%%%%%%%%%%%%%%%%%%%%
It should be mentioned that there exist different expressions of Bell's 
inequality. (\ref{ZZV_16}) is the original expression derived by Bell. 
Now we are able to verify its correctness for the QBB experiment. For 
this we consider the probabilities of the following 4 different 
experimental configurations:
%%%%%%%%%%%%%%%%%%%%%%%%%%%%%%%%%%%%%%%%%%%%%%%%%%%%%%%%%%%%%%%%%%%%%
\begin{eqnarray}
(\alpha,\beta) & = & \left(0,\frac{\pi}{8}\right)
\label{ZZV_17a}
\\
(\alpha,\beta') & = & \left(0,\frac{3 \pi}{8}\right)
\label{ZZV_17b}
\\
(\alpha',\beta) & = & \left(\frac{\pi}{4},\frac{\pi}{8}\right)
\label{ZZV_17c}
\\
(\alpha',\beta') & = & \left(\frac{\pi}{4},\frac{3 \pi}{8}\right) 
\; .
\label{ZZV_17d}
\end{eqnarray}
%%%%%%%%%%%%%%%%%%%%%%%%%%%%%%%%%%%%%%%%%%%%%%%%%%%%%%%%%%%%%%%%%%%%%
Surprisingly, Bell's inequality (\ref{ZZV_16}) is violated. A closer 
look onto the probabilities which belong to the different configurations 
reveals that configurations (\ref{ZZV_17a}), (\ref{ZZV_17c}), and 
(\ref{ZZV_17d}) result into the same probabilities, and, therefore, into 
identical correlation functions. This can also be inferred from the 
corresponding probability states. These are given by
\begin{itemize}
\item $\alpha=0$, $\beta=\pi/8$:
%%%%%%%%%%%%%%%%%%%%%%%%%%%%%%%%%%%%%%%%%%%%%%%%%%%%%%%%%%%%%%%%%%%%%
\begin{eqnarray}
|\Phi_{QBB}> \, =  \, \frac{1}{\sqrt 2} \cdot 
\sin(- \frac{\pi}{8}) \cdot |\varphi_1,\varphi_1> \, + ~~~~~~~~~~~~~
~~~~
\nonumber\\ 
\frac{1}{\sqrt 2} 
\cdot \cos(- \frac{\pi}{8}) \cdot |\varphi_1,\varphi_2> \, + 
~~~~~~~~~~ \nonumber\\
\frac{1}{\sqrt 2} \cdot \cos(- \frac{\pi}{8}) \cdot 
|\varphi_2,\varphi_1> \, 
+ \; \frac{1}{\sqrt 2} \cdot \sin(- \frac{\pi}{8}) \cdot 
|\varphi_2,\varphi_2>
\label{KMP_7a}
\end{eqnarray}
%%%%%%%%%%%%%%%%%%%%%%%%%%%%%%%%%%%%%%%%%%%%%%%%%%%%%%%%%%%%%%%%%%%%%
\item $\alpha=0$, $\beta=3\pi/8$:
%%%%%%%%%%%%%%%%%%%%%%%%%%%%%%%%%%%%%%%%%%%%%%%%%%%%%%%%%%%%%%%%%%%%%
\begin{eqnarray}
|\Phi_{QBB}> \, =  \, \frac{1}{\sqrt 2} \cdot 
\sin(- \frac{3 \pi}{8}) \cdot |\varphi_1,\varphi_1> \, + ~~~~~~~~~~~
~~~~~~
\nonumber\\ 
\frac{1}{\sqrt 2} \cdot \cos(- \frac{3 \pi}{8}) \cdot 
|\varphi_1,\varphi_2> \, +
~~~~~~~~~~~~~ \nonumber\\
\frac{1}{\sqrt 2} \cdot \cos(- \frac{3 \pi}{8}) \cdot 
|\varphi_2,\varphi_1> \, 
+ \; \frac{1}{\sqrt 2} \cdot \sin(- \frac{3 \pi}{8}) \cdot 
|\varphi_2,\varphi_2>
\label{KMP_7b}
\end{eqnarray}
%%%%%%%%%%%%%%%%%%%%%%%%%%%%%%%%%%%%%%%%%%%%%%%%%%%%%%%%%%%%%%%%%%%%%
\item $\alpha=\pi/4$, $\beta=\pi/8$:
%%%%%%%%%%%%%%%%%%%%%%%%%%%%%%%%%%%%%%%%%%%%%%%%%%%%%%%%%%%%%%%%%%%%%
\begin{eqnarray}
|\Phi_{QBB}> \, =  \, \frac{1}{\sqrt 2} \cdot 
\sin(\frac{\pi}{8}) \cdot |\varphi_1,\varphi_1> + \, \frac{1}{\sqrt 2} 
\cdot \cos(\frac{\pi}{8}) \cdot |\varphi_1,\varphi_2> \, + \nonumber
\\
\frac{1}{\sqrt 2} \cdot \cos(\frac{\pi}{8}) \cdot 
|\varphi_2,\varphi_1> \, 
+ \; \frac{1}{\sqrt 2} \cdot \sin(\frac{\pi}{8}) \cdot 
|\varphi_2,\varphi_2> ~~~~~~
\label{KMP_7c}
\end{eqnarray}
%%%%%%%%%%%%%%%%%%%%%%%%%%%%%%%%%%%%%%%%%%%%%%%%%%%%%%%%%%%%%%%%%%%%%
\item $\alpha=\pi/4$, $\beta=3\pi/8$:
%%%%%%%%%%%%%%%%%%%%%%%%%%%%%%%%%%%%%%%%%%%%%%%%%%%%%%%%%%%%%%%%%%%%%
\begin{eqnarray}
|\Phi_{QBB}> \, =  \, \frac{1}{\sqrt 2} \cdot 
\sin(- \frac{\pi}{8}) \cdot |\varphi_1,\varphi_1> \, + ~~~~~~~~~~~~
~~~~~
\nonumber\\ 
\frac{1}{\sqrt 2} 
\cdot \cos(- \frac{\pi}{8}) \cdot |\varphi_1,\varphi_2> \, + 
~~~~~~~~~~ \nonumber\\
\frac{1}{\sqrt 2} \cdot \cos(- \frac{\pi}{8}) \cdot 
|\varphi_2,\varphi_1> \, 
+ \; \frac{1}{\sqrt 2} \cdot \sin(- \frac{\pi}{8}) \cdot 
|\varphi_2,\varphi_2> \; .
\label{KMP_7d}
\end{eqnarray}
%%%%%%%%%%%%%%%%%%%%%%%%%%%%%%%%%%%%%%%%%%%%%%%%%%%%%%%%%%%%%%%%%%%%%
\end{itemize}
Thus we have in fact only 2 different experimental configurations. From 
(\ref{KMP_7a}) and (\ref{KMP_7b}) we can see moreover that the 
probabilities related to the single events are only interchanged.

\subsection{Description of Bell's experiment in terms of local\\
interactions}
\label{subsec:2}
%%%%%%%%%%%%%%%%%%%%%%%%%%%%%%%%%%%%%%%%%%%%%%%%%%%%%%%%%%%%%%%%%%%%%
\begin{figure}
\centering
\includegraphics[height=7.0cm]{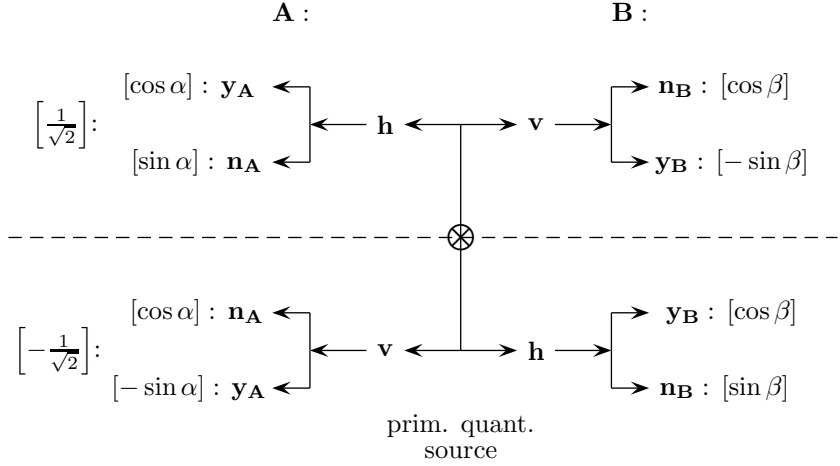}
\caption{General interaction scheme of Bell's experiments. $h$ and $v$ 
are the possible polarizations of the photons emitted by the primary 
source. $n_{A/B}$ (''lamp A/B off'') and $y_{A/B}$ (''lamp A/B on'') 
are the local events measured in the experiment after the additional 
local interactions took place. The corresponding probability 
amplitudes are given in the square brackets}
\label{Abbildung_4_3}
\end{figure}
%%%%%%%%%%%%%%%%%%%%%%%%%%%%%%%%%%%%%%%%%%%%%%%%%%%%%%%%%%%%%%%%%%%%
Fig. \ref{Abbildung_4_3} shows the general interaction scheme that holds 
for Bell's experiment (second level of configuration). $h$ represents a 
horizontally polarized photon emitted from the primary source into a 
certain direction. Correspondingly, $v$ represents a vertically 
polarized photon emitted by the same source and at the same time into 
the opposite direction. The local events measured on each side are 
$n_{A/B}$ (''lamp A/B off'') and $y_{A/B}$ (''lamp A/B on''). 
Eigenvector $|\varphi_1>$ is again related to $y_{A/B}$, and eigenvector 
$|\varphi_2>$ to $n_{A/B}$. We are now interested in a justification of 
the different probabilities of the 4 possible event pairs $(y_A,y_B), 
(y_A,n_B), (n_A,y_B)$, and $(n_A,n_B)$ measured before in the QBB 
experiment. The 2 substates before the interaction, which are the 
result of the primary source, are given by 
%%%%%%%%%%%%%%%%%%%%%%%%%%%%%%%%%%%%%%%%%%%%%%%%%%%%%%%%%%%%%%%%%%%%%
\begin{equation}
|\Phi^{(0)}_1> \, =  \, \frac{1}{\sqrt{2}} \cdot 
|\varphi_1,\varphi_2>
\label{QKBK_1a}
\end{equation}
%%%%%%%%%%%%%%%%%%%%%%%%%%%%%%%%%%%%%%%%%%%%%%%%%%%%%%%%%%%%%%%%%%%%%
and
%%%%%%%%%%%%%%%%%%%%%%%%%%%%%%%%%%%%%%%%%%%%%%%%%%%%%%%%%%%%%%%%%%%%%
\begin{equation}
|\Phi^{(0)}_2> \, =  \, - \, \frac{1}{\sqrt{2}} \cdot 
|\varphi_2,\varphi_1> \; .
\label{QKBK_1b}
\end{equation}
%%%%%%%%%%%%%%%%%%%%%%%%%%%%%%%%%%%%%%%%%%%%%%%%%%%%%%%%%%%%%%%%%%%%%
They belong obviously to orthogonal subspaces and can therefore be 
superposed without resulting in an interference term.

The 2 local stochastic interactions on each side and the resulting local
probability amplitudes (the sine and cosine functions in the square 
brackets in Fig \ref{Abbildung_4_3}) may be obtained by use of local  
and unitary T-matrices which are identical with the matrix of rotation,
%%%%%%%%%%%%%%%%%%%%%%%%%%%%%%%%%%%%%%%%%%%%%%%%%%%%%%%%%%%%%%%%%%%%%
\begin{equation}
{\bf T_{\alpha/\beta}} \, = \, {\bf D_{\alpha/\beta}} \, = \, 
\left( \begin{array}{cc}
{\cos\alpha/\beta} & {- \sin\alpha/\beta}\\
{\sin\alpha/\beta} & {\cos\alpha/\beta}
\end{array} \right) \; .
\label{Kap2_16a}
\end{equation}
%%%%%%%%%%%%%%%%%%%%%%%%%%%%%%%%%%%%%%%%%%%%%%%%%%%%%%%%%%%%%%%%%%%%%
The amplitudes $c_{1'}$ and $c_{2'}$ of the local states 
after the interaction are then the result of relation
%%%%%%%%%%%%%%%%%%%%%%%%%%%%%%%%%%%%%%%%%%%%%%%%%%%%%%%%%%%%%%%%%%%%%
\begin{equation}
 \left( \begin{array}{c}
{c_{1'}}\\
{c_{2'}}
\end{array} \right) \, = \, {\bf T_{\alpha/\beta}} \cdot 
\left( \begin{array}{c}
{c_1}\\
{c_2}
\end{array} \right)
\label{Kap2_31a}
\end{equation}
%%%%%%%%%%%%%%%%%%%%%%%%%%%%%%%%%%%%%%%%%%%%%%%%%%%%%%%%%%%%%%%%%%%%%
with $c_1$ and $c_2$ being the amplitudes  of the local 
states before the interaction. This description is identical with the 
description of the classical interaction of a linearly polarized plane 
wave with a polarization filter rotated by an angle of $\alpha$ (or 
$\beta$) against the plane of linear polarization. Such T-matrices are 
the decisive elements in electromagnetic wave scattering as well as in 
quantum scattering theory (see \cite{book2,book3}, for example). After 
the interaction, from (\ref{Kap2_31a}) and the respective T-matrices we 
get therefore the following local probability states of the 4 2-dim. 
subspaces:
\begin{itemize}
\item local observation point $A$ and $h$-polarization:
%%%%%%%%%%%%%%%%%%%%%%%%%%%%%%%%%%%%%%%%%%%%%%%%%%%%%%%%%%%%%%%%%%%%%
\begin{equation}
|\phi(A,h)> \, =  \, \frac{1}{\sqrt{2}} \cdot \left( 
\cos\alpha \cdot |\varphi_1> \, + \; \sin\alpha \cdot |\varphi_2>
\right)
\label{pbell_1a}
\end{equation}
%%%%%%%%%%%%%%%%%%%%%%%%%%%%%%%%%%%%%%%%%%%%%%%%%%%%%%%%%%%%%%%%%%%%%
\item local observation point $A$ and $v$-polarization:
%%%%%%%%%%%%%%%%%%%%%%%%%%%%%%%%%%%%%%%%%%%%%%%%%%%%%%%%%%%%%%%%%%%%%
\begin{equation}
|\phi(A,v)> \, =  \, - \, \frac{1}{\sqrt{2}} \cdot \left( - \, 
\sin\alpha \cdot |\varphi_1> \, + \; \cos\alpha \cdot |\varphi_2>
\right)
\label{pbell_1b}
\end{equation}
%%%%%%%%%%%%%%%%%%%%%%%%%%%%%%%%%%%%%%%%%%%%%%%%%%%%%%%%%%%%%%%%%%%%%
\item local observation point $B$ and $h$-polarization:
%%%%%%%%%%%%%%%%%%%%%%%%%%%%%%%%%%%%%%%%%%%%%%%%%%%%%%%%%%%%%%%%%%%%%
\begin{equation}
|\phi(B,h)> \, =  \, \cos\beta \cdot |\varphi_1> \, + \; 
\sin\beta \cdot |\varphi_2>
\label{pbell_1c}
\end{equation}
%%%%%%%%%%%%%%%%%%%%%%%%%%%%%%%%%%%%%%%%%%%%%%%%%%%%%%%%%%%%%%%%%%%%%
\item local observation point $B$ and $v$-polarization:
%%%%%%%%%%%%%%%%%%%%%%%%%%%%%%%%%%%%%%%%%%%%%%%%%%%%%%%%%%%%%%%%%%%%%
\begin{equation}
|\phi(B,v)> \, =  \, - \, \sin\beta \cdot |\varphi_1> \, + \; 
\cos\beta \cdot |\varphi_2> \; .
\label{pbell_1d}
\end{equation}
%%%%%%%%%%%%%%%%%%%%%%%%%%%%%%%%%%%%%%%%%%%%%%%%%%%%%%%%%%%%%%%%%%%%%
\end{itemize}
Please, note that we have assigned the probability amplitudes of the 
primary source to the local probability states on the left hand side. 
In the next step we pass on from these local substates to the direct 
product states which belong to the event pairs from both sides. This 
procedure is accomplished separately for the upper and lower part 
of Fig. \ref{Abbildung_4_3}. That's because these parts belong to 
different experimental situations. Moreover, performing the direct 
product requires that the local interactions on each side are 
stochastically independent. In doing so, we get the following 2 
probability states:
%%%%%%%%%%%%%%%%%%%%%%%%%%%%%%%%%%%%%%%%%%%%%%%%%%%%%%%%%%%%%%%%%%%%%
\begin{eqnarray}
|\Phi_1> \, = \, |\phi(A,h)> \, \otimes \; 
|\phi(B,v)> \; = ~~~~~~~~~~~~~~~~~~~~~~~~~
\nonumber\\ 
\frac{1}{\sqrt{2}} \cdot \left[ \, - \, \cos\alpha 
\cdot \sin\beta \cdot |\varphi_1,\varphi_1> \, + \; \cos\alpha \cdot 
\cos\beta \cdot |\varphi_1,\varphi_2> \, - \right.
\nonumber\\
\left. \sin\alpha \cdot \sin\beta \cdot |\varphi_2,\varphi_1> \, + \;
\sin\alpha \cdot \cos\beta \cdot |\varphi_2,\varphi_2> \, \right] 
~~~~
\label{QKBK_2a}
\end{eqnarray}
%%%%%%%%%%%%%%%%%%%%%%%%%%%%%%%%%%%%%%%%%%%%%%%%%%%%%%%%%%%%%%%%%%%%%
(this state belongs to the upper part of Fig. \ref{Abbildung_4_3}), 
and
%%%%%%%%%%%%%%%%%%%%%%%%%%%%%%%%%%%%%%%%%%%%%%%%%%%%%%%%%%%%%%%%%%%%%
\begin{eqnarray}
|\Phi_2>  \, = \, |\phi(A,v)> \, \otimes \; 
|\phi(B,h)> \; = ~~~~~~~~~~~~~~~~~~~~~~~~~
\nonumber\\
\frac{1}{\sqrt{2}} \cdot \left[ \, \sin\alpha 
\cdot \cos\beta \cdot |\varphi_1,\varphi_1> \, + \; \sin\alpha \cdot 
\sin\beta \cdot |\varphi_1,\varphi_2> \, - \right.
\nonumber\\
\left. \cos\alpha \cdot \cos\beta \cdot |\varphi_2,\varphi_1> \, - \;
\cos\alpha \cdot \sin\beta \cdot |\varphi_2,\varphi_2> \, \right] 
~~~~ 
\label{QKBK_2b}
\end{eqnarray}
%%%%%%%%%%%%%%%%%%%%%%%%%%%%%%%%%%%%%%%%%%%%%%%%%%%%%%%%%%%%%%%%%%%%%
(this state belongs to the lower part of Fig. \ref{Abbildung_4_3}).
These 2 probability states are obviously {\bf not entangled!} And, 
although they are orthogonal, they {\bf do not belong to orthogonal 
subspaces any longer!} Now it is quite easy to see that the 
superposition of these both non-entangled substates results indeed 
into state (\ref{ZZV_6}) with amplitudes (\ref{ZZV_7}) and (\ref{ZZV_8}) 
(except that now $c_{21} = - c_{12}$ holds, as already mentioned).

Introducing the normalized probability states
%%%%%%%%%%%%%%%%%%%%%%%%%%%%%%%%%%%%%%%%%%%%%%%%%%%%%%%%%%%%%%%%%%%%%
\begin{eqnarray}
|{\tilde \Phi}_1> \, := \, - \, \cos\alpha \cdot \sin\beta 
\cdot |\varphi_1,\varphi_1> \, + \, \cos\alpha \cdot \cos\beta \cdot 
|\varphi_1,\varphi_2> 
\, - 
\nonumber\\
\sin\alpha \cdot \sin\beta \cdot |\varphi_2,\varphi_1> \, + \,
\sin\alpha \cdot \cos\beta \cdot |\varphi_2,\varphi_2>
\label{pbell_4a}
\\
|{\tilde \Phi}_2> \, := \, \sin\alpha \cdot \cos\beta \cdot 
|\varphi_1,\varphi_1> \, + \, \sin\alpha \cdot \sin\beta \cdot 
|\varphi_1,\varphi_2> \, - ~~~~
\nonumber\\
\cos\alpha \cdot \cos\beta \cdot |\varphi_2,\varphi_1> \, - \,
\cos\alpha \cdot \sin\beta \cdot |\varphi_2,\varphi_2> 
\label{pbell_4b}
\\
|{\tilde \Phi}_3> \, := \, \frac{1}{\sqrt{2}} \cdot \left[ 
|\varphi_1,\varphi_2> \, + \; |\varphi_2,\varphi_1>\right] 
~~~~~~~~~~~~~~~~~~~~~~~~~~~~~~~~
\label{pbell_4c}
\\
|{\tilde \Phi}_4> \, := \, \frac{1}{\sqrt{2}} \cdot \left[
|\varphi_1,\varphi_1> \, + \; |\varphi_2,\varphi_2>\right] 
~~~~~~~~~~~~~~~~~~~~~~~~~~~~~~~~
\label{pbell_4d}
\end{eqnarray}
%%%%%%%%%%%%%%%%%%%%%%%%%%%%%%%%%%%%%%%%%%%%%%%%%%%%%%%%%%%%%%%%%%%%%
will allow us to relate a statistical operator to the quantum mechanical 
Bell's experiment. This operator reads
%%%%%%%%%%%%%%%%%%%%%%%%%%%%%%%%%%%%%%%%%%%%%%%%%%%%%%%%%%%%%%%%%%%%%
\begin{equation}
{\bf {\hat \rho}^{\, (QBB)}} \, := \, \sum_{i=1}^4 \, 
p_i \cdot |{\tilde \Phi}_i><{\tilde \Phi}_i| \; ,
\label{pbell_5}
\end{equation}
%%%%%%%%%%%%%%%%%%%%%%%%%%%%%%%%%%%%%%%%%%%%%%%%%%%%%%%%%%%%%%%%%%%%%
with the weights given by
%%%%%%%%%%%%%%%%%%%%%%%%%%%%%%%%%%%%%%%%%%%%%%%%%%%%%%%%%%%%%%%%%%%%%
\begin{eqnarray}
p_1 & = & p_2 \, = \, \frac{1}{2}
\label{pbell_6a}
\\
p_3 & = & - \, p_4 \, = \, 2 \cdot c(\alpha,\beta) \; ,
\label{pbell_6b}
\end{eqnarray}
%%%%%%%%%%%%%%%%%%%%%%%%%%%%%%%%%%%%%%%%%%%%%%%%%%%%%%%%%%%%%%%%%%%%%
and
%%%%%%%%%%%%%%%%%%%%%%%%%%%%%%%%%%%%%%%%%%%%%%%%%%%%%%%%%%%%%%%%%%%%%
\begin{equation}
c(\alpha,\beta) \, = \, \sin\alpha \cdot \sin\beta \cdot 
\cos\alpha \cdot \cos\beta \; .
\label{pbell_3}
\end{equation}
%%%%%%%%%%%%%%%%%%%%%%%%%%%%%%%%%%%%%%%%%%%%%%%%%%%%%%%%%%%%%%%%%%%%%
$|{\tilde \Phi}_1>$ and $|{\tilde \Phi}_2>$ as well as $|{\tilde \Phi}_3>$ 
and $|{\tilde \Phi}_4>$ are again orthogonal among each other. Moreover, 
$|{\tilde \Phi}_3>$ and $|{\tilde \Phi}_4>$ belong to orthogonal subspaces, 
in contrast to $|{\tilde \Phi}_1>$ and $|{\tilde \Phi}_2>$. Please, note 
also the negative weight (negative quasi probability) $p_4$! The measured 
probabilities (\ref{ZZV_10}) and (\ref{ZZV_11}) are then the result of
%%%%%%%%%%%%%%%%%%%%%%%%%%%%%%%%%%%%%%%%%%%%%%%%%%%%%%%%%%%%%%%%%%%%%
\begin{eqnarray}
w(y,y) & = & <\varphi_1,\varphi_1| {\bf {\hat \rho}^{\, (QBB)}} 
|\varphi_1,\varphi_1>
\label{QKBK_3a}
\\
w(n,n) & = & <\varphi_2,\varphi_2| {\bf {\hat \rho}^{\, (QBB)}} 
|\varphi_2,\varphi_2>
\label{QKBK_3b}
\\
w(y,n) & = & <\varphi_2,\varphi_1| {\bf {\hat \rho}^{\, (QBB)}} 
|\varphi_1,\varphi_2>
\label{QKBK_3c}
\\
w(n,y) & = & <\varphi_1,\varphi_2| {\bf {\hat \rho}^{\, (QBB)}} 
|\varphi_2,\varphi_1> \; ,
\label{QKBK_3d}
\end{eqnarray}
%%%%%%%%%%%%%%%%%%%%%%%%%%%%%%%%%%%%%%%%%%%%%%%%%%%%%%%%%%%%%%%%%%%%%
as usually known from quantum mechanics. Operator (\ref{pbell_5}) may 
be called the ''basic Bell's operator'' since it is related to a single 
experiment with a fixed parameter configuration $\alpha$ and $\beta$. 
On the other hand, if we have a mixture of $N$ such experiments (for 
different parameter configurations $(\alpha_k,\beta_k)$ with $k=1, 
\cdots ,N$) with the classical weights $r_k$, $\sum_{k=1}^N r_k=1$, 
then we have the following statistical operator of the mixture:
%%%%%%%%%%%%%%%%%%%%%%%%%%%%%%%%%%%%%%%%%%%%%%%%%%%%%%%%%%%%%%%%%%%%%
\begin{equation}
{\bf {\hat R}} \, = \, \sum_{k=1}^N \, r_k \cdot 
{\bf {\hat \rho}_k^{\, (QBB)}} \, = \, \sum_{k=1}^N \, \sum_{i=1}^4 
\, r_k \cdot p_i^{(k)} \cdot 
|{\tilde \Phi}^{(k)}_i><{\tilde \Phi}^{(k)}_i| \; .
\label{Gemisch_1}
\end{equation}
%%%%%%%%%%%%%%%%%%%%%%%%%%%%%%%%%%%%%%%%%%%%%%%%%%%%%%%%%%%%%%%%%%%%%
This is again well-known from quantum mechanics where similar operators 
are used to describe incoherent mixtures of pure quantum states.

Regarding the statistical operator (\ref{pbell_5}) the question of the 
linear independence of the normalized probability states 
$|{\tilde \Phi}_i>$ is of some importance. If this happens, then we are 
able to represent any probability state of our 4-dim. event space by 
a linear combination of these state vectors. To prove the linear 
independence we have to consider Grams' matrix
%%%%%%%%%%%%%%%%%%%%%%%%%%%%%%%%%%%%%%%%%%%%%%%%%%%%%%%%%%%%%%%%%%%%%
\begin{equation}
{\bf G} \, = \, <{\tilde \Phi}_i|{\tilde \Phi}_j> \; ; \quad 
i,j \, = \, 1,2 \; .
\label{Gram_1}
\end{equation}
%%%%%%%%%%%%%%%%%%%%%%%%%%%%%%%%%%%%%%%%%%%%%%%%%%%%%%%%%%%%%%%%%%%%%
Because of (\ref{pbell_4a}) - (\ref{pbell_4d}) this matrix is a
symmetric ones,
%%%%%%%%%%%%%%%%%%%%%%%%%%%%%%%%%%%%%%%%%%%%%%%%%%%%%%%%%%%%%%%%%%%%%
\begin{eqnarray}
{\bf G} \, = \, \left( \begin{array}{cccc}
{1} & {0} & {g_1} & {g_2}\\
{0} & {1} & {-g_1} & {g_2}\\
{g_1} & {-g_1} & {1} & {0}\\
{g_2} & {g_2} & {0} & {1} \end{array} \right) \; ,
\label{Gram_2}
\end{eqnarray}
%%%%%%%%%%%%%%%%%%%%%%%%%%%%%%%%%%%%%%%%%%%%%%%%%%%%%%%%%%%%%%%%%%%%%
with elements $g_1$ and $g_2$ given by
%%%%%%%%%%%%%%%%%%%%%%%%%%%%%%%%%%%%%%%%%%%%%%%%%%%%%%%%%%%%%%%%%%%%%
\begin{eqnarray}
g_1 & = & \frac{1}{\sqrt{2}} \cdot \cos(\alpha \, + \, \beta)
\label{Gram_2a}
\\
g_2 & = &  \frac{1}{\sqrt{2}} \cdot \sin(\alpha \, - \, \beta) \; .
\label{Gram_2b}
\end{eqnarray}
%%%%%%%%%%%%%%%%%%%%%%%%%%%%%%%%%%%%%%%%%%%%%%%%%%%%%%%%%%%%%%%%%%%%%
Its determinant reads
%%%%%%%%%%%%%%%%%%%%%%%%%%%%%%%%%%%%%%%%%%%%%%%%%%%%%%%%%%%%%%%%%%%%%
\begin{equation}
\mbox{det}\left({\bf G}\right) \, = \, \frac{8}{\sqrt{2}} \cdot 
c(\alpha,\beta) \; ,
\label{Gram_3}
\end{equation}
%%%%%%%%%%%%%%%%%%%%%%%%%%%%%%%%%%%%%%%%%%%%%%%%%%%%%%%%%%%%%%%%%%%%%
with $c(\alpha,\beta)$ according to (\ref{pbell_3}). Thus we have to 
meet the condition 
%%%%%%%%%%%%%%%%%%%%%%%%%%%%%%%%%%%%%%%%%%%%%%%%%%%%%%%%%%%%%%%%%%%%%
\begin{equation}
c(\alpha,\beta) \, \ne \, 0 
\label{Gram_4}
\end{equation}
%%%%%%%%%%%%%%%%%%%%%%%%%%%%%%%%%%%%%%%%%%%%%%%%%%%%%%%%%%%%%%%%%%%%%
to ensure the linear independence of the vectors (\ref{pbell_4a}) - 
(\ref{pbell_4d}). On the other hand, if
%%%%%%%%%%%%%%%%%%%%%%%%%%%%%%%%%%%%%%%%%%%%%%%%%%%%%%%%%%%%%%%%%%%%%
\begin{equation}
c(\alpha,\beta) \, = \, 0 \; ,
\label{Gram_4a}
\end{equation}
%%%%%%%%%%%%%%%%%%%%%%%%%%%%%%%%%%%%%%%%%%%%%%%%%%%%%%%%%%%%%%%%%%%%%
holds (this happens if we have $(\alpha=0 \; \mbox{or} \; \pi/2,\beta 
\ne 0)$ or $(\alpha \ne 0,\beta=0 \; \mbox{or} \; \pi/2)$), both 
weights $p_3$ and $p_4$ are identical zero. In other words: The 
probabilities of the events represented by the state vectors 
$|{\tilde \Phi}_3>$ (a ''source'' of the sum of probabilities 
$w(y,n) + w(n,y)$) and $|{\tilde \Phi}_4>$ (a ''sink'' of the sum of 
probabilities $w(y,y) + w(n,n)$) of the 2 orthogonal subspaces will 
no longer be redistributed to the probabilities of the events 
represented by the state vectors $|{\tilde \Phi}_1>$ and 
$|{\tilde \Phi}_2>$. 

The essential result of the above considerations is the fact that the 
interference term, that results from the superposition of the two 
non-entangled probability states (\ref{QKBK_2a}) and (\ref{QKBK_2b}), 
is responsible for the violation of Bell's inequality. This interference 
term was traced back to additional local and stochastically independent 
interactions. Thus, there seems to be no need to assume any ''spooky 
action at a distance'' or any ''hidden parameters'' behind the 
probability states to understand the quantum mechanical Bell's 
experiment. But the following aspect is also of some importance. As one 
can see from Eqs. (\ref{ZZV_14}) and (\ref{pbell_6b})/(\ref{pbell_3}), 
entanglement is not necessarily linked to the existence of negative quasi 
probabilities. However, such statements can frequently be found in the 
literature (see \cite{journal8}, for example). If we choose 
$(\alpha=\pi/8,\beta=0)$ as the local interaction parameters, for example, 
then the quasi probabilities $p_3$ and $p_4$ are identical zero. But the 
corresponding probability state of the QBB is still entangled! The 
existence of negative quasi probabilities is therefore neither a 
sufficient nor a necessary condition of entanglement.

\subsection{Description of Bell's experiment in terms of a basis\\
transformation}
\label{subsec:3}
The quantum mechanical Bell's experiment has been described so far in 
terms of local interactions accomplished by 2 additional polarization 
filters. However, in several presentations I was faced again and again 
with an incomprehension regarding this interaction point of view. 
Therefore, let's discuss the quantum mechanical Bell's experiment again 
but from the point of view of the well accepted basis transformation to 
demonstrate the equivalence of both approaches. 

We ask for the transformation matrix that transforms the primary 
probability state 
%%%%%%%%%%%%%%%%%%%%%%%%%%%%%%%%%%%%%%%%%%%%%%%%%%%%%%%%%%%%%%%%%%%%%
\begin{equation}
|\Phi^{(0)}> \, =  \, \frac{1}{\sqrt{2}} \cdot \left[
|\varphi_1,\varphi_2> \, - \; |\varphi_2,\varphi_1> \right]
\label{WWBell_1}
\end{equation}
%%%%%%%%%%%%%%%%%%%%%%%%%%%%%%%%%%%%%%%%%%%%%%%%%%%%%%%%%%%%%%%%%%%%%
of the QBB into the new probability state
%%%%%%%%%%%%%%%%%%%%%%%%%%%%%%%%%%%%%%%%%%%%%%%%%%%%%%%%%%%%%%%%%%%%%
\begin{eqnarray}
|\Phi_{QBB}> \, =  \, c_{yy} \cdot |y_A,y_B> + \, c_{yn} 
\cdot |y_A,n_B> + ~~~~~~~~~~~~~~~~~~~
\nonumber\\
\, c_{ny} \cdot |n_A,y_B> + \, c_{nn} \cdot 
|n_A,n_B>
\label{WWBell_2}
\end{eqnarray}
%%%%%%%%%%%%%%%%%%%%%%%%%%%%%%%%%%%%%%%%%%%%%%%%%%%%%%%%%%%%%%%%%%%%%
with probability amplitudes
%%%%%%%%%%%%%%%%%%%%%%%%%%%%%%%%%%%%%%%%%%%%%%%%%%%%%%%%%%%%%%%%%%%%%
\begin{eqnarray}
c_{yy} & = & c_{nn} \, = \, \frac{1}{\sqrt 2} \cdot 
\sin (\alpha - \beta)
\label{WWBell_2a}
\\
c_{yn} & = & - \, c_{ny} \, = \, \frac{1}{\sqrt 2} \cdot 
\cos (\alpha - \beta) \; .
\label{WWBell_2b}
\end{eqnarray}
%%%%%%%%%%%%%%%%%%%%%%%%%%%%%%%%%%%%%%%%%%%%%%%%%%%%%%%%%%%%%%%%%%%%%
The new but so far unknown eigenvectors $|y_{A/B}>$ und $|n_{A/B}>$ 
are again related to the possible local measurements ''lamp A/B on'' 
and ''lamp A/B off''. Now, let's assume that these eigenvectors are 
the result of a rotation of the local coordinate system on the left- 
and right hand side, 
%%%%%%%%%%%%%%%%%%%%%%%%%%%%%%%%%%%%%%%%%%%%%%%%%%%%%%%%%%%%%%%%%%%%%
\begin{equation}
 \left( \begin{array}{c}
{|y_{A/B}>}\\
{|n_{A/B}>}
\end{array} \right) \, = \, {\bf D_{\alpha/\beta}} \cdot 
\left( \begin{array}{c}
{|\varphi_1>}\\
{|\varphi_2>}
\end{array} \right) \; ,
\label{WWBell_3}
\end{equation}
%%%%%%%%%%%%%%%%%%%%%%%%%%%%%%%%%%%%%%%%%%%%%%%%%%%%%%%%%%%%%%%%%%%%%
caused by the 2 polarization filters. ${\bf D_{\alpha/\beta}}$ therein 
is again the matrix (\ref{Kap2_16a}) of rotation. Thus we get
%%%%%%%%%%%%%%%%%%%%%%%%%%%%%%%%%%%%%%%%%%%%%%%%%%%%%%%%%%%%%%%%%%%%%
\begin{eqnarray}
|y_A> & = & (\cos\alpha, - \sin\alpha)
\label{WWBell_4a}
\\
|y_B> & = & (\cos\beta, - \sin\beta) \; .
\label{WWBell_4b}
\\
|n_A> & = & (\sin\alpha,\cos\alpha)
\label{WWBell_4c}
\\
|n_B> & = & (\sin\beta,\cos\beta) \; .
\label{WWBell_4d}
\end{eqnarray}
%%%%%%%%%%%%%%%%%%%%%%%%%%%%%%%%%%%%%%%%%%%%%%%%%%%%%%%%%%%%%%%%%%%%%
Next we introduce the T-matrix
%%%%%%%%%%%%%%%%%%%%%%%%%%%%%%%%%%%%%%%%%%%%%%%%%%%%%%%%%%%%%%%%%%%%%
\begin{eqnarray}
{\bf T} \, = \, \left( \begin{array}{cc}
{<y_B,y_A|\varphi_1,\varphi_2>} & {<y_B,y_A|\varphi_2,\varphi_1>}\\
{<n_B,y_A|\varphi_1,\varphi_2>} & {<n_B,y_A|\varphi_2,\varphi_1>}\\
{<y_B,n_A|\varphi_1,\varphi_2>} & {<y_B,n_A|\varphi_2,\varphi_1>}\\
{<n_B,n_A|\varphi_1,\varphi_2>} & {<n_B,n_A|\varphi_2,\varphi_1>}
\end{array} \right) \, = ~~~~~~~~~ \nonumber\\
\left( \begin{array}{cc}
{-\cos\alpha \cdot \sin\beta} & {-\sin\alpha \cdot \cos\beta}\\
{\cos\alpha \cdot \cos\beta} & {-\sin\alpha \cdot \sin\beta}\\
{-\sin\alpha \cdot \sin\beta} & {\cos\alpha \cdot \cos\beta}\\
{\sin\alpha \cdot \cos\beta} & {\cos\alpha \cdot \sin\beta}
\end{array} \right)
\label{WWBell_5}
\end{eqnarray}
%%%%%%%%%%%%%%%%%%%%%%%%%%%%%%%%%%%%%%%%%%%%%%%%%%%%%%%%%%%%%%%%%%%%%
Then the new probability amplitudes (\ref{WWBell_2a}) and 
(\ref{WWBell_2b}) are calculated according to
%%%%%%%%%%%%%%%%%%%%%%%%%%%%%%%%%%%%%%%%%%%%%%%%%%%%%%%%%%%%%%%%%%%%%
\begin{equation}
\left( \begin{array}{c}
{c_{yy}}\\
{c_{yn}}\\
{c_{ny}}\\
{c_{nn}}
\end{array} \right) \, = \, {\bf T} \cdot 
\left( \begin{array}{c}
{1/\sqrt{2}}\\
{-1/\sqrt{2}}
\end{array} \right) 
\label{WWBell_6}
\end{equation}
%%%%%%%%%%%%%%%%%%%%%%%%%%%%%%%%%%%%%%%%%%%%%%%%%%%%%%%%%%%%%%%%%%%%%
from the primary probabiliy amplitudes of (\ref{WWBell_1}). The T-matrix 
(\ref{WWBell_5}) is again a unitary matrix, i.e.,
%%%%%%%%%%%%%%%%%%%%%%%%%%%%%%%%%%%%%%%%%%%%%%%%%%%%%%%%%%%%%%%%%%%%%
\begin{equation}
{\bf T}^{\, tp} \cdot {\bf T} \, = \, {\bf E}
\label{WWBell_7}
\end{equation}
%%%%%%%%%%%%%%%%%%%%%%%%%%%%%%%%%%%%%%%%%%%%%%%%%%%%%%%%%%%%%%%%%%%%%
holds with ${\bf E}$ representing the $2 \times 2$ unit matrix. This 
ensures the conservation of the total probability. However, it is a 
disadvantage of the basis transformation that the aspect of the 
superposition of 2 non-entangled substates is covered up. But a closer 
look onto (\ref{WWBell_5}) reveals that the probability amplitudes of 
these 2 substates are identical with the elements of the first and 
second column of (\ref{WWBell_5}). The fact that T-matrices can be 
used for both the description of a certain interaction and the 
description of an equivalent transformation of corresponding 
eigenvectors is considered in detail in \cite{book2}.

\section{A classical Bell's experiment}
\label{sec:3}
In this section we will discuss the classical counterpart of the 
quantum mechanical Bell's experiment. To start with let's venture the 
following guess: Regarding the statistical operator (\ref{pbell_5}) 
we expect that non-vanishing weights $p_3$ and $p_4$ are the essential 
aspect that makes the quantum mechanical Bell's experiment differ from 
its classical counterpart. If this is true, then we would get the 
probabilities of a corresponding classical experiment by use of 
(\ref{QKBK_3a}) - (\ref{QKBK_3d}) but with 
${\bf {\hat \rho}^{\, (QBB)}}$ replaced by the classical statistical 
operator
%%%%%%%%%%%%%%%%%%%%%%%%%%%%%%%%%%%%%%%%%%%%%%%%%%%%%%%%%%%%%%%%%%%%%
\begin{equation}
{\bf {\hat \rho}^{\, (cl.)}} \, := \, \sum_{i=1}^2 \, 
p_i \cdot |{\tilde \Phi}_i><{\tilde \Phi}_i| \; .
\label{pbell_5mod}
\end{equation}
%%%%%%%%%%%%%%%%%%%%%%%%%%%%%%%%%%%%%%%%%%%%%%%%%%%%%%%%%%%%%%%%%%%%%
This corresponds to the procedure that we first calculate the 
probabilities of each substate (\ref{QKBK_2a}) and (\ref{QKBK_2b}) 
separately, and if adding up these probabilities afterwards. In this 
way we would end up with the following classical probabilities: 
\begin{itemize}
\item Probability $w(y,y)$/$w(n,n)$ that both lamps are switched 
on/switched off:
%%%%%%%%%%%%%%%%%%%%%%%%%%%%%%%%%%%%%%%%%%%%%%%%%%%%%%%%%%%%%%%%%%%%%
\begin{equation}
w(y,y) \, = \, w(n,n) \, =  \, \frac{1}{2} \cdot \left( \sin^2\alpha 
\cdot \cos^2\beta \, + \, \sin^2\beta \cdot \cos^2\alpha \right)
\label{ZZV_10mod}
\end{equation}
%%%%%%%%%%%%%%%%%%%%%%%%%%%%%%%%%%%%%%%%%%%%%%%%%%%%%%%%%%%%%%%%%%%%%
\item Probability $w(y,n)$/$w(n,y)$ that just one lamp is switched 
on and the other lamp remains switched off:
%%%%%%%%%%%%%%%%%%%%%%%%%%%%%%%%%%%%%%%%%%%%%%%%%%%%%%%%%%%%%%%%%%%%%
\begin{equation}
w(n,y) \, = \, w(y,n) \, =  \, \frac{1}{2} \cdot \left( \cos^2\alpha 
\cdot \cos^2\beta \, + \, \sin^2\beta \cdot \sin^2\alpha \right) \; .
\label{ZZV_11mod}
\end{equation}
%%%%%%%%%%%%%%%%%%%%%%%%%%%%%%%%%%%%%%%%%%%%%%%%%%%%%%%%%%%%%%%%%%%%%
\end{itemize}
It would be of some benefit to prove these probabilities in a 
corresponding experiment, as it was done in the quantum mechanical 
case by A. Aspect and co-workers. And there is indeed a quite simple 
marble experiment that can be used to prove the correctness of the 
probabilities (\ref{ZZV_10mod}) and (\ref{ZZV_11mod}). This experiment 
is described in what follows for the 2 parameter configurations 
(\ref{ZZV_17a}) and (\ref{ZZV_17b}), i.e. the only 2 configurations of 
the 4 different QBB experiments required to verify Bell's inequality 
which result in different correlation functions. But the extension to 
other parameter configurations is straightforward.
\newline

\noindent
{\it A Box $B_p$ with 1 white and 1 black marble represents the primary 
stochastic source. 2 additional boxes $B_w$ and $B_b$ are filled with 
17 white and 3 black marbles (box $B_w$) and 17 black and 3 white marbles 
(box $B_b$). This corresponds approximately to the probabilities of 
$0.85$/$0.15$ to draw a white or black marble out of the respective box. 
These 2 additional boxes represent the local interaction on the 
right hand side! 

Now, if the parameter configuration (\ref{ZZV_17a}) is chosen, the 
experiment runs as follows:
We draw both marbles blindly out of box $B_p$ and put one marble on the 
left hand side and the other marble on the right hand side on our desk. 
The colour of the marble on the left hand side is already the result of this 
side. To get the result of the right hand side requires an additional step. 
If the primary marble on the right hand side is white, then we have to draw 
another marble out of box $B_w$. Its colour is the result of the right hand 
side. But if the primary marble on the right hand side is black, then we have 
to draw another marble out of box $B_b$. This colour will then be the result 
of the right hand side. We repeat this procedure until we are able to 
calculate the probabilities within a sufficient accuracy. Then we are able 
to calculate the correlation function $K(\alpha=0,\beta=\pi/8)$.

If the parameter configuration (\ref{ZZV_17b}) is chosen, the 
experiment runs as follows: The first step to get the result of the left 
hand side is as before. But, now, if the primary marble on the right hand 
side is white, then we have to draw another marble out of box $B_b$. Its 
colour is the result of the right hand side. On the other hand, if the 
primary marble on the right hand side is black, then we have to draw another 
marble out of box $B_w$. This colour will then be the result of the right 
hand side. We repeat this procedure until we are able to calculate the 
probabilities within a sufficient accuracy. Then we are able to calculate 
the correlation function $K(\alpha=0,\beta=3\pi/8)$.

We can proceed in a similar way if the local parameter $\alpha$ is not zero. 
The only thing we have to do is to fill 2 additional boxes on the left hand 
side with an appropriate number of black and white marbles to meet the 
probabilities of the local interaction on this side. At least 
200 single experiments for each parameter configuration are needed to 
approach the probabilities (\ref{ZZV_10mod}) and (\ref{ZZV_11mod}) within a 
sufficient accuracy.}
\newline

\noindent
Once we accept the classical probabilities (\ref{ZZV_10mod}) and 
(\ref{ZZV_11mod}) as an experimental fact, and if we compare these 
probabilities with the quantum mechanical probabilities (\ref{ZZV_10}) 
and (\ref{ZZV_11}) we are able to make the following statements:
\begin{itemize}
\item Performing the classical Bell's experiment with the 4 different 
sets (\ref{ZZV_17a}) - (\ref{ZZV_17d}) of the local interaction 
parameters shows that a violation of Bell's inequality (\ref{ZZV_16}) 
cannot be observed! Thus we may state that Bell's inequality can be used 
to distinguish whether Bell's experiment was performed with classical or 
quantum mechanical objects.
\item The probabilities of the quantum mechanical and classical Bell's 
experiment are identical if $(\alpha=0,\beta \ne 0)$, or 
if $(\alpha \ne 0,\beta=0)$. From this we may 
conclude that:
\item The probabilities of the quantum mechanical Bell's experiment with 
fixed parameters $\alpha_{QBB}$ and $\beta_{QBB}$ can be reproduced by a 
corresponding classical Bell's experiment with fixed parameters $(\alpha_{cl}=0,\beta_{cl}=\alpha_{QBB}-\beta_{QBB})$ or $(\beta_{cl}=0,\alpha_{cl}=\alpha_{QBB}-\beta_{QBB})$. Thus, looking 
only at the probabilities of a certain experiment without having the 
information about the local interaction parameters will not allow us 
to distinguish if this experiment was performed with classical or quantum 
objects.
\item Each local observer of a Bell's experiment will always measure the 
probabilities $1/2$ for both local events ''lamp on: $y_{A/B}$'' or ''lamp 
of: $n_{A/B}$'', independent of the local interaction parameters $\alpha$ 
and $\beta$, and independent of whether classical or quantum objects are 
used.
\item Beside the violation of Bell's inequality there exists another 
criterion to decide whether Bell's experiment was performed with quantum 
mechanical or classical objects. In contrast to the 4 necessary experimental 
configurations required to test Bell's inequality we can simply choose 
one experimental configuration with the same local interaction parameters, 
let's say $\alpha=\beta=\pi/4$, for example. If the quantum mechanical 
Bell's experiment was performed, then the probabilities for the 4 events 
are given by $w(y,y)=w(n,n)=0$ and $w(y,n)=w(n,y)=1/2$. On the other hand, 
we get the probabilities $w(y,y)=w(n,n)=w(y,n)=w(n,y)=1/4$ if the classical 
Bell's experiment was performed.
\end{itemize}

\section{Bell's experiment and complementarity}
\label{sec:4}
This final section is concerned with the following question: What would be 
the result of the quantum mechanical Bell's experiment if performing 
noncontacting measurements to detect the polarization state of the photon 
pairs emitted by the primary source, and before the additional local 
interactions with the polarization filters on both sides take place? It is 
the same situation we are faced with in the quantum mechanical double slit 
experiment. There we may ask what happens with the characteristic frequency 
distribution in the far field if performing noncontacting measurements to 
get the ''which slit''-information. Both questions are strongly related to 
the complementarity principle. According to the usual understanding of the 
complementarity principle, and regarding the double slit experiment we would 
expect that the characteristic interference pattern will be lost if getting 
the ''which slit''-information by noncontacting measurements. But the very 
recent experiments with polarization entangled photons and a special laser 
mode performed at the University of Potsdam have shown that the 
characteristic double slit interference pattern can still be observed (see 
\cite{journal7,journal10}). The ''which slit''-information was obtained 
by noncontacting coincidence measurements in these experiments. I.e., 
despite the fact that the ''which slit''-information was obtained both 
substates (the states related to each of the slits) must be superposed to 
explain this behaviour. If this result is transfered to the quantum 
mechanical Bell's experiment the situation becomes as follows:

Let's assume that we are equipped with a noncontacting measurement setup to 
get the information about the state of polarization of the primary emitted 
photon pairs, and before the additional local interactions with the 
polarization filters on each side of Bell's experiment take place. According 
to the usual understanding of the complementarity principle one may expect 
that the probabilities (\ref{ZZV_10mod}) and (\ref{ZZV_11mod}) would be 
measured, instead of the probabilities (\ref{ZZV_10}) and (\ref{ZZV_11}). 
That's because we are then in exactly the same situation we are confronted 
with in the classical marble experiment. Knowing the state of polarization 
of a photon would allow us to predict the probability of the photon to pass 
a polarization filter in a certain orientation. Therefore, the superposition 
of both substates (\ref{QKBK_2a}) and (\ref{QKBK_2b}) should be excluded. 
Such a noncontacting measurement of the state of polarization of a photon 
pair would be possible if there exists a primary source that emits 3 or more 
polarization entangled photons at the same time. The additional entangled 
photons may be used for coincidence measurements, for example, as it was done 
in the Potsdam experiments. Unfortunately, no such source or experiment is 
described in the literature, so far. But if we transfer the result of the 
Potsdam double slit experiment to the quantum mechanical Bell's experiment 
we may speculate that we would still measure the probabilities (\ref{ZZV_10}) 
and (\ref{ZZV_11}). However, there is a quite interesting indication that 
supports our speculation, and that holds for the Potsdam double slit experiment 
as well. If we would indeed measure the probabilities (\ref{ZZV_10mod}) and 
(\ref{ZZV_11mod}) if performing the noncontacting measurements to get the 
''which state of polarization'' information, then probabilities would no longer 
represent objectively measurable quantities since depending on the state of 
information of the observer! Two observers, one equipped with a noncontacting 
measurement setup and the other without such an equipment, would measure 
different probabilities in the same experiment. This is not even conceivable 
since the considered event space is a pure classical one. The result of a 
measurement should be factual and independent of any state of information of 
the observer. Otherwise, physics would lose its objective character since 
objective measurements are its backbone. Abandon the objectivity of 
measurements would have serious consequences not only for quantum mechanics 
but for general physics. Fortunately, the result of the Potsdam experiment, 
and the speculative result of Bell's experiment are (or would be) in 
agreement with the requirement of the objectivity of any but especially 
quantum mechanical measurements.

The above discussion raises the question if we have to distinguish carefully 
between ''objective measurements'' and ''noncontacting measurements'' (or, 
better, the noncontacting gain of information). The former 
should be factual and independent of any state of information of the observer. 
Contrary, the latter is based on a certain knowledge (or theory) gained by 
performing local or non-local but objective measurements before. This 
knowledge is used afterwards in future experiments. This seems 
to be the only way to perform ''noncontacting measurements'', in my opinion. 
And that's what was exactly done in the Potsdam experiments. In contrast to 
''noncontacting measurements'' the ''objective measurements'' are based 
essentially on the interaction of the object under consideration with the 
measurement device. This interaction has an uncontrollable impact on the 
state of the object itself, as well-known from quantum mechanics. Therefore, 
if the Potsdam experiments would not be performed with noncontacting but 
objective measurements on the signal photon to get the ''which slit'' 
information, then we may expect an impact on the characteristic interference 
pattern in the far field.

\section{Conclusion}
\label{sec:5}
The most essential results of this paper may be condensed into the 
following 2 statements:
\begin{enumerate}
\item Entanglement is not responsible for the violation of Bell's 
inequality in the quantum mechanical Bell's experiment. It is in fact 
the superposition of 2 orthogonal and non-entangled substates, as 
already known from the quantum mechanical double slit experiment. 
The two substates which must be superposed in the quantum mechanical case 
turned out to be the result of 2 additional, {\bf local, and stochastically 
independent interactions.}
\item It is possible to introduce a statistical operator for both the 
quantum mechanical and classical Bell's experiment. This operator 
contains already negative quasi probabilities in the quantum mechanical case. 
But these negative quasi probabilities are neither sufficient nor necessary 
for entantglement. They are rather an indication of the existence of 
interference terms.
\end{enumerate}
But what makes the superposition of probability states in quantum mechanics 
staying in conflict with our everyday experience? It cannot be the 
superposition itself since this process is already known from classical 
field theories if there exist 2 or more fields at the same time in a certain 
region of space. To answer the question we must first note that the measurable 
events in Bell's as well as in the quantum mechanical double slit experiment 
are typical particlelike (the flashing of a pixel on a screen or a lamp, a 
click, etc.). All these events are local with respect to time and position and 
are usually attributed to the interaction of a particle with the measurement 
device. The very notation ''photon'' provokes already the picture of a particle, 
irregardless of the abstract formalism behind. 
Looking at the superposition of the 2 substates in the quantum mechanical Bell's 
experiment we have to note that both substates belong to events which do not 
take place at the same time. But they must be superposed, as though existing at 
the same time, to get the correct probabilities. The same holds for the quantum 
mechanical doube slit experiment. That's indeed strange and hard to understand. 
Could this problem possibly be solved by reconsidering our understanding of 
''simultaneity'' on the atomic and subatomic level? That's what happened already 
on a macroscopic scale in the context of special 
relativity. The point of view that ''probability'' should usually represent a 
quantity that is independent of time (comparable to the steady state situation 
of classical fields?) would be an argument against this idea. Moreover, in 
quantum mechanics probability is rather related to the behaviour of an ensemble 
than to that of a single particle. And if an ensemble is considered in 
the equilibrium state we have the equality of time- and ensemble average. 
However, this aspect of time seems to be worth to further considerations, to my 
mind.


\section*{References}
\begin{thebibliography}{10}

\bibitem{journal1} Einstein A., Podolsky B., Rosen, N.: ''Can Quantum-Mechanical 
Description of Physical Reality be Considered Complete?'', Phys. Rev. 38, 777-780 
(1935)
\bibitem{journal2} Bohr, N.: ''Can Quantum-Mechanical Description of Physical 
Reality be Considered Complete?'', Phys. Rev. 38, 696-702 (1935)
\bibitem{book1} Bell, J. S.: ''Speakable and unspeakable in quantum mechanics'', 
Univ. Press, Cambridge (1993) 
\bibitem{journal3} Aspect, A., Grangier, P., Roger, G.: ''Experimental Tests of 
Bell's Inequalities Using Time-Varying Analyzers'', Phys. Rev. Let.49, 1804-1807 
(1982)
\bibitem{journal4} Schroedinger, E.: ''Discussion of Probability Relations Between 
Separated Systems'', Cambridge Phil. Soc. 31, 555-563 (1935)
\bibitem{journal5} Shimony, A.: ''Bell's Theorem'', web address:
\newline
\noindent ''http://plato.stanford.edu/entries/bell-theorem/''
\bibitem{book2} Rother, T.: ''Electromagnetic Wave Scattering on 
Nonspherical Particles: Basic Methodology and Simulations'', Springer, New 
York Heidelberg Berlin (2009)
\bibitem{book3} Newton, R. G.: ''Scattering Theory of Waves and Particles'', 
Springer, New York Heidelberg Berlin (1982)
\bibitem{journal8} Sperling, J., Vogel, W.: ''Representation of entanglement by 
negative quasi-probabilities'', Phys. Rev. A {\bf 79}, 042337 (2009)
\bibitem{journal7} Menzel, R., et al.: ''A two-photon double-slit experiment'', 
J. Mod. Optics, 1 - 9, DOI: 10.1080/09500340.2012.746400 (2012)
\bibitem{journal10} Menzel, R., et al.: ''Wave-particle dualism and complementarity 
unraveled by a different mode''. PNAS {\bf 109}, S. 9314 - 9319 (2012)

\end{thebibliography}
\end{document}